\documentclass[
  journal=pasa,
  manuscript=research-paper, 
  year=2025,
  volume=YY,
]{cup-journal}

\usepackage{amsmath}
\usepackage{tablefootnote}
\usepackage{hyperref}
\usepackage{amssymb,microtype,siunitx,booktabs}
\usepackage{academicons}
\usepackage{xcolor}
\usepackage{longtable}
\usepackage{supertabular}
\usepackage{multicol}
\usepackage{siunitx}
\usepackage{tabularx}

\title{EMU/GAMA: Refining Dust Extinction Corrections for H$\alpha$ Luminosity Functions Using Radio-Based Calibration}

\usepackage{orcidlink}

\author{J. Willingham\textsuperscript{\orcidlink{0000-0002-0545-1113}}} 
\affiliation{School of Mathematical and Physical Sciences, 12 Wally's Walk, Macquarie University, NSW 2109, Australia}
\email[J. Willingham]{jayde.willingham@hdr.mq.edu.au}
\alsoaffiliation{Macquarie University Astrophysics and Space Technologies Research Centre, Sydney, NSW 2109, Australia}

\author{A. Hopkins\textsuperscript{\orcidlink{0000-0002-6097-2747}}}
\affiliation{School of Mathematical and Physical Sciences, 12 Wally's Walk, Macquarie University, NSW 2109, Australia}
\alsoaffiliation{Macquarie University Astrophysics and Space Technologies Research Centre, Sydney, NSW 2109, Australia}

\author{T. Zafar\textsuperscript{\orcidlink{0000-0003-3935-7018}}}
\affiliation{School of Mathematical and Physical Sciences, 12 Wally's Walk, Macquarie University, NSW 2109, Australia}
\alsoaffiliation{Macquarie University Astrophysics and Space Technologies Research Centre, Sydney, NSW 2109, Australia}
\alsoaffiliation{ARC Centre of Excellence for All Sky Astrophysics in 3 Dimensions (ASTRO-3D), Australia}

\author{J. Afonso \textsuperscript{\orcidlink{0000-0002-9149-2973}}}
\affiliation{Instituto de Astrofísica e Ciências do Espaço, Universidade de Lisboa, OAL, Tapada da Ajuda, 1349-018 Lisbon, Portugal}
\alsoaffiliation{Departamento de Física, Faculdade de Ciências, Universidade de Lisboa, Edifício C8, Campo Grande, 1749-016 Lisbon, Portugal}

\author{U.T. Ahmed\textsuperscript{\orcidlink{0000-0002-0309-1599}}}
\affiliation{Australian Astronomical Optics, Macquarie University, 105 Delhi Rd, North Ryde, NSW 2113, Australia}
\alsoaffiliation{Macquarie University Astrophysics and Space Technologies Research Centre, Sydney, NSW 2109, Australia}
\alsoaffiliation{Centre for Astrophysics, University of Southern Queensland, 37 Sinnathamby Boulevard, Springfield Central, QLD 4300, Australia}

\author{A. Ahmad\textsuperscript{\orcidlink{0000-0002-0457-3661}}}
\affiliation{School of Science, Western Sydney University, Locked Bag 1797, Penrith, NSW 2751, Australia}

\author{A. Battisti\textsuperscript{\orcidlink{0000-0003-4569-2285}}}
\affiliation{Research School of Astronomy and Astrophysics, Australian National University, Weston Creek, ACT, Australia}
\alsoaffiliation{International Centre for Radio Astronomy Research (ICRAR), University of Western Australia, M468, Crawley, WA, Australia}
\alsoaffiliation{ARC Centre of Excellence for All Sky Astrophysics in 3 Dimensions (ASTRO-3D), Australia}

\author{D. Bomans\textsuperscript{\orcidlink{0000-0001-5126-5365}}}
\affiliation{Ruhr Astroparticle and Plasma Physics Center (RAPP Center), 44780 Bochum, Germany}
\alsoaffiliation{Ruhr-Universität Bochum, Fakultät für Physik und Astronomie, Astronomisches Institut (AIRUB), 44780 Bochum, Germany}

\author{M. J. I. Brown\textsuperscript{\orcidlink{0000-0002-1207-9137}}}
\affiliation{School of Physics \& Astronomy, Monash University, Clayton, VIC, Australia}

\author{M. Cowley\textsuperscript{\orcidlink{0000-0002-4653-8637}}}
\affiliation{School of Chemistry \& Physics, Faculty of Science, Queensland University of Technology, Brisbane, QLD 4000, Australia}
\alsoaffiliation{University of Southern Queensland, Centre for Astrophysics, West Street, Toowoomba QLD 4350, Australia}

\author{D. Farrah\textsuperscript{\orcidlink{0000-0003-1748-2010}}}
\affiliation{Department of Physics and Astronomy, University of Hawai'i at Mānoa, 2505 Correa Road, Honolulu, HI 96822, USA}
\alsoaffiliation{Institute for Astronomy, University of Hawai'i, 2680 Woodlawn Dr., Honolulu, HI 96822, USA}

\author{T.J. Galvin\textsuperscript{\orcidlink{0000-0002-2801-766X}}}
\affiliation{International Centre for Radio Astronomy Research, Curtin University, Bentley, WA 6102, Australia}
\alsoaffiliation{CSIRO Space \& Astronomy, P.O. Box 1130, Bentley, WA 6102, Australia}

\author{B. Holwerda\textsuperscript{\orcidlink{0000-0002-4884-6756}}}
\affiliation{Leiden Observatory, Leiden University, NL-2300 RA Leiden, Netherlands}

\author{D. Leahy\textsuperscript{\orcidlink{0000-0002-4814-958X}}}
\affiliation{Department of Physics and Astronomy, University of Calgary, Calgary, AB, Canada}

\author{U. Maio\textsuperscript{\orcidlink{0000-0002-0039-3102}}}
\affiliation{INAF – Osservatorio Astronomico di Trieste, Via Tiepolo 11, I-34143, Trieste, Italy}
\alsoaffiliation{IFPU – Institute for Fundamental Physics of the Universe, Via Beirut 2, I-34014, Trieste, Italy}

\author{T. Mukherjee\textsuperscript{\orcidlink{0009-0004-7639-869X}}}
\affiliation{School of Mathematical and Physical Sciences, 12 Wally's Walk, Macquarie University, NSW 2109, Australia}
\alsoaffiliation{Macquarie University Astrophysics and Space Technologies Research Centre, Sydney, NSW 2109, Australia}
\alsoaffiliation{ARC Centre of Excellence for All Sky Astrophysics in 3 Dimensions (ASTRO-3D), Australia}

\author{J. Prathap\textsuperscript{\orcidlink{0009-0004-0251-2672}}}
\affiliation{School of Mathematical and Physical Sciences, 12 Wally's Walk, Macquarie University, NSW 2109, Australia}
\alsoaffiliation{Macquarie University Astrophysics and Space Technologies Research Centre, Sydney, NSW 2109, Australia}
\alsoaffiliation{ARC Centre of Excellence for All Sky Astrophysics in 3 Dimensions (ASTRO-3D), Australia}

\author{N. Seymour\textsuperscript{\orcidlink{0000-0003-3506-5536}}}
\affiliation{International Centre for Radio Astronomy Research, Curtin University, GPO Box U1987, Bentley, WA 6845, Australia}

\author{J.Th. van Loon\textsuperscript{\orcidlink{0000-0002-1272-3017}}}
\affiliation{Lennard-Jones Laboratories, Keele University, ST5 5BG, UK}

\author{E. Ward\textsuperscript{\orcidlink{0009-0007-3696-5153}}}
\affiliation{School of Mathematical and Physical Sciences, 12 Wally's Walk, Macquarie University, NSW 2109, Australia}
\alsoaffiliation{Macquarie University Astrophysics and Space Technologies Research Centre, Sydney, NSW 2109, Australia}

\received {dd Mmm 2025}
\revised  {dd Mmm 2025}
\accepted {dd Mmm 2025}
\published{dd Mmm 2025}


\begin{document}

\begin{abstract}
  We present a novel approach to correcting H$\alpha$ luminosity functions for dust extinction by calibrating against radio-based star formation rates (SFRs), using data from the Evolutionary Map of the Universe (EMU) and Galaxy and Mass Assembly (GAMA) surveys. Accurate dust correction is essential for deriving SFRs from rest-frame UV-optical emission lines, particularly as the \textit{James Webb Space Telescope} extends such measurements to galaxies at $z>5$. While a luminosity dependence of dust obscuration has long been recognised, our method exploits the empirical relationship between obscured (H$\alpha$) and unobscured (radio) SFRs to provide a dust correction that can be applied where traditional spectroscopic techniques, e.g. Balmer line based approaches, are unavailable. We apply the SFR based dust correction to 25 published H$\alpha$ luminosity functions spanning $0<z<8$, and derive corresponding star formation rate densities (SFRDs). Adopting the locally calibrated H$\alpha$--radio relation ends up with an overestimate of the cosmic SFRD by more than two orders of magnitude at $z\gtrsim1$. Motivated by the luminosity dependent relation in the local Universe, we introduce a new model where the luminosity dependence of the dust obscuration decreases with increasing redshift. This approach can reproduce observed SFRDs across cosmic time. These results highlight the potential of a radio-based  calibration for dust correction, where a luminosity dependent correction would need to decline in strength with increasing redshift. This implies that the dust content or distribution in galaxies at early epochs differs substantially from that in the local Universe.  
\end{abstract}

\section{Introduction }
\label{sec:int}

Galaxies are a key building block of the Universe as the primary source of photons at low to high redshifts. Understanding the physical processes that govern their formation and transformation is therefore essential to modern astrophysics and understanding the Universe. 
As a fundamental quantity in extragalactic astronomy, the star formation rate (SFR) serves as a powerful diagnostic of galactic activity and offers a direct probe of how galaxies build up their stellar mass and evolve over cosmic time \citep{kennicutt_star_1998, madau_cosmic_2014, madau_radiation_2017}.

SFR is not a quantity that can be measured directly in distant galaxies where individual stars are no longer resolved. Instead, it is inferred from the luminosity of a galaxy at specific wavelengths that trace recent star-forming activity or inferred from spectral energy distribution modelling across a range of wavelengths. These galaxies are typically identified by their strong ultraviolet (UV) emission from young, massive stars and infrared (IR) emission resulting from the re-radiation of starlight by interstellar dust \citep{kennicutt_star_1998}. The most commonly used approaches for estimating SFR consider UV,  H$\alpha$, IR, and radio wavelength calibrations. While no single method is ideal, extensive research has been devoted to comparing, combining, and correcting these different tracers to improve consistency and accuracy \citep{hopkins_toward_2001, bell_estimating_2003, hopkins_evolution_2004, calzetti_calibration_2007, kennicutt_dust-corrected_2009, hao_dust-corrected_2011, davies_gamah-atlas_2016, davies_galaxy_2017, brown_calibration_2017}.

A powerful and widely used SFR tracer in the optical regime is the H$\alpha$ emission line. This line arises from recombination in HII regions that form when UV photons from high mass, short-lived O and B stars ionise surrounding hydrogen gas. Because these stars have lifespans of $\lesssim10$~Myr, the H$\alpha$ emission line directly traces recent and ongoing star formation \citep{glazebrook_measurement_1999, kennicutt_star_2012}.
Like other optical tracers, H$\alpha$ emission is subject to dust attenuation, which necessitates corrections. Accurate measurement of H$\alpha$ flux also requires accounting for underlying stellar absorption, where the absorption features from older stellar populations partially fill in the emission line. Correcting for dust attenuation effects depends on the quality of the stellar continuum modelling and subtraction, which can be challenging, particularly in large spectroscopic surveys.

One of the most widely used tools to infer dust attenuation is the Balmer Decrement, BD=F(H$\alpha$)/F(H$\beta$), which provides a basis for correcting dust-affected emission line fluxes. Balmer recombination lines are produced in HII regions by young, massive stars \citep{osterbrock_astrophysics_1989}. The BD method relies on comparing the observed ratio of Balmer line fluxes (typically H$\alpha$/H$\beta$) to the theoretically predicted case B recombination value, with any excess reddening interpreted as dust attenuation. By adopting an extinction law, the wavelength-dependent obscuration can then be quantified and applied to correct emission-line–derived star formation rates.

A limitation of the BD method is that it typically assumes a uniform dust screen geometry and treats the galaxy as optically thin. In reality, galaxies often exhibit complex, clumpy dust distributions and optically thick central regions, which may lead to underestimates of the true extinction \citep{calzetti_dust_2001, kreckel_mapping_2013, robertson_ground-_2024}. Reliable measurements require high signal-to-noise ratios for both the H$\alpha$ and H$\beta$ emission lines, which can be challenging, especially for high-redshift or faint galaxy systems. Low S/N in faint emission lines can also introduce significant biases in line-ratio diagnostics \citep{yuan_systematics_2013}. At higher redshifts, detections of both H$\alpha$ and H$\beta$ remain uncommon, with surveys such as HETDEX reporting only three confirmed LAEs at 
$z\gtrsim2$ (of which only two have multiple rest-frame optical emission lines detected) \citep{finkelstein_hetdex_2011}.

Radio wavelengths are particularly advantageous in star formation studies due to their ability to penetrate the dust that often obscures ultraviolet and optical tracers. Though radio emission is not a direct tracer of star formation activity, its utility as a SFR indicator stems from the well-established correlation between radio and far infrared (FIR) luminosities in star-forming galaxies \citep{condon_radio_1992, yun_radio_2001, bell_estimating_2003, ivison_far-infraredradio_2010, molnar_non-linear_2021}. In addition, radio emission has also been shown to correlate well with dust-corrected H$\alpha$ luminosities \citep[e.g.,][]{murphy_calibrating_2011}, further reinforcing its connection to recent massive star formation and providing a useful bridge between radio and traditional recombination-line tracers. Previous studies have also directly calibrated H$\alpha$ against FIR emission at $z\sim1-2$, providing important cross-checks on dust-corrected SFR estimates \citep[e.g.,][]{sobral_large_2013, ibar_herschel_2013}. These comparisons highlight both the potential and limitations of using FIR as a reference, motivating the need for refined dust corrections when relying on H$\alpha$ at higher redshift.

Radio emission has traditionally been studied at $1.4$\,GHz, though with the advent of many Square Kilometre Array pathfinder telescopes and programs, this is rapidly expanding to much a broader range of frequencies. Radio emission at frequencies lower than a few GHz is dominated by non-thermal synchrotron radiation produced by cosmic-ray electrons accelerated in supernova remnants. This is an indirect tracer of recent massive star formation over timescales of $\sim$10–100\,Myr \citep{murphy_calibrating_2011}. At higher frequencies, thermal free-free emission contributes an increasing fraction of the total radio continuum. This emission originates from ionised gas in H II regions and directly traces massive star formation, analogous to H$\alpha$ or H$\beta$. Around $\sim$30 GHz, the free-free and synchrotron components become comparable in strength \citep{condon_radio_1992}.

Contamination from AGN can complicate interpretations, necessitating careful source classification \citep{cowley_zfourge_2016}. Once dust extinction is accounted for, H$\alpha$ has relatively few assumptions for the conversion to star formation rate (i.e. recent star formation history and initial mass function) and thus is less susceptible to systematic errors than some tracers.
The relative insensitivity of radio emission to dust obscuration raises the possibility that statistical measurements of star formation in the radio could be used to inform or calibrate dust corrections for optical tracers like H$\alpha$, but this approach has not been comprehensively explored.

The aim of this study is to develop a new approach for dust obscuration correction in measurements of star formation, particularly at high redshifts where traditional methods become increasingly challenging \citep{sanders_aurora_2024}. Dust correction is likely the single dominant source of uncertainty at intermediate redshifts $1 < z< 5$, which means an empirical SFR based dust correction method, that is dependent on luminosity and redshift, could have great impact in this epoch. This need is motivated by the frequent absence of wavelength coverage and/or sufficient signal-to-noise to perform Balmer decrement–based corrections for individual galaxies in current deep surveys \citep[e.g.,][]{reddy_paschen-line_2023, matharu_first_2023, shapley_jwstnirspec_2023}. To address this limitation, we exploit the well-established relationship between H$\alpha$ and radio tracers in the local Universe (e.g., \citet{hopkins_toward_2001}, Ahmed et al., submitted). 
This link provides a pathway to extend dust-correction techniques to the early Universe, where the \textit{James Webb Space Telescope} (JWST) has significantly expanded our ability to probe star-forming galaxies, enabling detailed observations of systems out to redshifts $z > 10$ \citep{pontoppidan_jwst_2022}. Such surveys offer a vital dataset for studying the evolution of star formation through H$\alpha$ emission at the earliest epochs, provided that robust obscuration corrections can be established. To this end, we adopt a model for luminosity dependent obscuration anchored in the local relation, and progressively reduce its strength with redshift, using the cosmic star formation history as a constraint. This approach allows us to infer the likely redshift dependence of luminosity dependent obscuration in H$\alpha$, thereby enabling physically motivated corrections applicable to the earliest galaxies.

The layout of this paper is as follows. In \S\ref{sec:Data} the radio and optical data used in order to fit a SFR tracer relationship is introduced. In this same section a compilation of published H$\alpha$ luminosity function results are presented, along with details on dust correction removal and IMF conversion where needed. Then in \S\ref{Dust correction} the novel approach to dust correction for H$\alpha$ luminosities is introduced and applied to published work. The results of this new dust correction are fitted for their luminosity function (LF) and star formation rate density (SFRD) evolution in \S\ref{Results}. 
Throughout we assume cosmological parameters of $H_0=70\,$km\,s$^{-1}$\,Mpc$^{-1}$, $\Omega_M=0.3$,
$\Omega_\Lambda=0.7$ and $\Omega_{\rm{k}} = 0$, a convenient approximation to recent measurements \citep[e.g.,][]{planck_collaboration_planck_2016}. Adopting more precise cosmological values does not change any of our analysis measurably. We adopt a Chabrier IMF, converting other published results to this IMF as needed.

\section{Data}
\label{sec:Data}

This analysis requires data from multiple catalogues to create a multiwavelength study of star forming galaxies (SFGs) that connects the obscured H$\alpha$ SFR tracer to the unobscured radio SFR tracer, for the same set of sources over a large enough coverage for a statistical study. We use early science radio data from the Evolutionary Map of the Universe \citep[EMU,][]{2011PASA...28..215N, 2021PASA...38...46N,hopkins_evolutionary_2025}, focusing on the G23 field. This dataset is integrated with far-ultraviolet (FUV) and far-infrared (FIR) data from the Galaxy and Mass Assembly (GAMA) survey \citep{2009A&G....50e..12D, 10.1111/j.1365-2966.2010.18188.x, 10.1093/mnras/stac472, 2015MNRAS.452.2087L}. The combined EMU-GAMA data, within a redshift range of $0.0<z<0.35$\footnote{The redshift range is limited by the ability to detect the H$\alpha$ line in the wavelength range of the AAOmega spectrograph on the AAT \citep{10.1111/j.1365-2966.2010.18188.x}.}, is used to compare uncorrected H$\alpha$-based SFRs to radio-derived (1.4\,GHz) SFRs, which serve as a dust-unbiased reference. Given that the radio luminosity is unaffected by dust obscuration we adopt it as a proxy for the corrected SFR measure for H$\alpha$. 
This analysis is then used to correct, or recorrect, for dust obscuration in other published H$\alpha$ luminosity function studies spanning the last $\sim$30\,yrs.

\subsection{Evolutionary Map of the Universe}
The Evolutionary Map of the Universe (EMU) is an ongoing radio continuum survey \citep{hopkins_evolutionary_2025} aimed at providing a comprehensive view of the Southern sky using the Australian Square Kilometre Array Pathfinder (ASKAP) telescope \citep{2007PASA...24..174J, McConnell_al._2020}. ASKAP employs an array of $36\times12\,$m antennas, covering baselines from 22 to 6000\,m. This configuration enables observations in the 800 – 1800\,MHz frequency range with an instantaneous bandwidth of 288\,MHz. EMU's use of ASKAP is expected to catalogue approximately 20 million galaxies, providing the most comprehensive atlas of the southern sky, with an angular resolution of $\sim 15''$ FWHM and a sensitivity of roughly $20\,\mu$Jy beam$^{-1}$.

For this study, we use EMU early science observations from the GAMA G23 field, \citep{gurkan_deep_2022}; see also \citet{Leahy_al._2019} for earlier technical details. The G23 region, centered at $\alpha = 23$\,h and $\delta = -32^\circ$, spans an area of $82.7\,$deg$^2$, with observations at a frequency of 887.5~MHz and a sensitivity of 0.038 $\mu$Jy beam$^{-1}$. The G23 radio data are matched with the GAMA catalogue to support our analysis.

\subsection{Galaxy and Mass Assembly}

The GAMA survey comprises photometric and spectroscopic data across five sky fields (G02, G09, G12, G15, and G23), covering a total area of approximately 280\,deg$^2$. Spectroscopic data were obtained using the AAOmega spectrograph on the Anglo-Australian Telescope (AAT), which supports a maximum of 400 fibres and spectral resolution of up to 10,000 \citep{2004SPIE.5492..410S, Sharp_2006}. Of particular interest for this study is the G23 field that covers 82.7\,deg$^2$, ($339^\circ<$RA$<351^\circ$, $-35^\circ<\delta<-30^\circ$),  with a limiting magnitude of $i<19.2$\,mag at a frequency of 887.5\,MHz and was targeted for early science observations from EMU \cite{gurkan_deep_2022}.

We use the latest data from the GAMA Data Release~4 (DR4) in this analysis, accessing three primary catalogues stored in data management units (DMUs) \citep{bellstedt_galaxy_2020, 10.1093/mnras/stac472}. The first is the \texttt{StellarMassesv24} table, which includes data on stellar masses, rest-frame photometry, and population synthesis fits across the five survey regions \citep{2011MNRAS.418.1587T}. 
Photometric data for the G23 field are obtained from the \texttt{gkvInputCatv02} DMU, which integrates data from the European Southern Observatory’s (ESO) Visible and Infrared Survey Telescope for Astronomy (VISTA) Kilo-Degree Infrared Galaxy Public Survey \citep[VIKING;][]{2013Msngr.154...32E} and the ESO VST Kilo Degree Survey \citep[KiDS;][]{jong_first_2015}. GKV (GAMA-KiDS-VIKING), provides multiwavelength photometry spanning from the far ultraviolet, using the Galaxy Evolution Explorer (GALEX) space telescope \citep[100–200 nm;][]{2015MNRAS.452.2087L}, to the second WISE band \citep[W2; 4.6\,$\mu$m;][]{2010AJ....140.1868W}. This DMU is not directly accessed here but is integral to related and ongoing studies of this dataset, including applications such as spectral energy distribution fitting.
Emission line fluxes and equivalent width measurements are accessed from the \texttt{GaussFitSimplev05} DMU \citep{2017MNRAS.465.2671G}. This catalogue employs multiple Gaussian fits to enhance accuracy. Previous studies have shown that simpler models yield comparable quantitative results \citep{Ahmed_al._2024}.

\subsection{Sample Selection}
This study focuses on comparing SFR estimators that draw on a diverse set of parameters. To achieve this, GAMA data was cross-matched with that from EMU within the G23 field.
In order to achieve this multiwavelength approach, several GAMA catalogues are combined and identical sources were identified based on either spectral or catalogue IDs, leading to a total of 25,762 GAMA sources with emission line and photometric data. A 5-arcsecond cross-match with the EMU catalogue, motivated by the cross-match success curves from \citet{Ahmed_al._2024}, yielded a ``parent sample'' of 6,633 sources. 

The parent sample was refined using several quality assurance markers, resulting in a final sample of 1,036 sources. A detailed explanation of the criteria used is presented here.

The first quality assurance criterion pertains to the Balmer lines, H$\alpha$ and H$\beta$, as well as the NII and OIII emission lines used in the BPT diagnostic. Here a positive detection threshold was established to ensure accurate calculations of luminosities and, consequently, SFRs.
Additionally, these emission line fluxes were required to meet a signal-to-noise (S/N) ratio threshold of $\rm{S/N}>3$. These cuts are the deepest impact to the sample size but it should guarantee that this work is built on a quality dataset.

Redshift is necessary for calculating luminosities. Redshift quality is assessed using the parameter \( nQ \geq 3 \) which corresponds to a redshift of at least 90\% confidence  \citep{10.1093/mnras/stv1436}. After applying this criterion, relevant equivalent widths were tested for positivity, and a ceiling was applied to the emission line fluxes.

Then a flag within the GAMA \texttt{GaussFitSimple} catalogue indicates whether the spectrum used provides the best redshift fit, as some objects have several spectra. The IS\_BEST condition is included to further enhance the robustness of the sample. 

\begin{figure}
    \centering
    \includegraphics[width=\linewidth,trim=20 25 5 5]{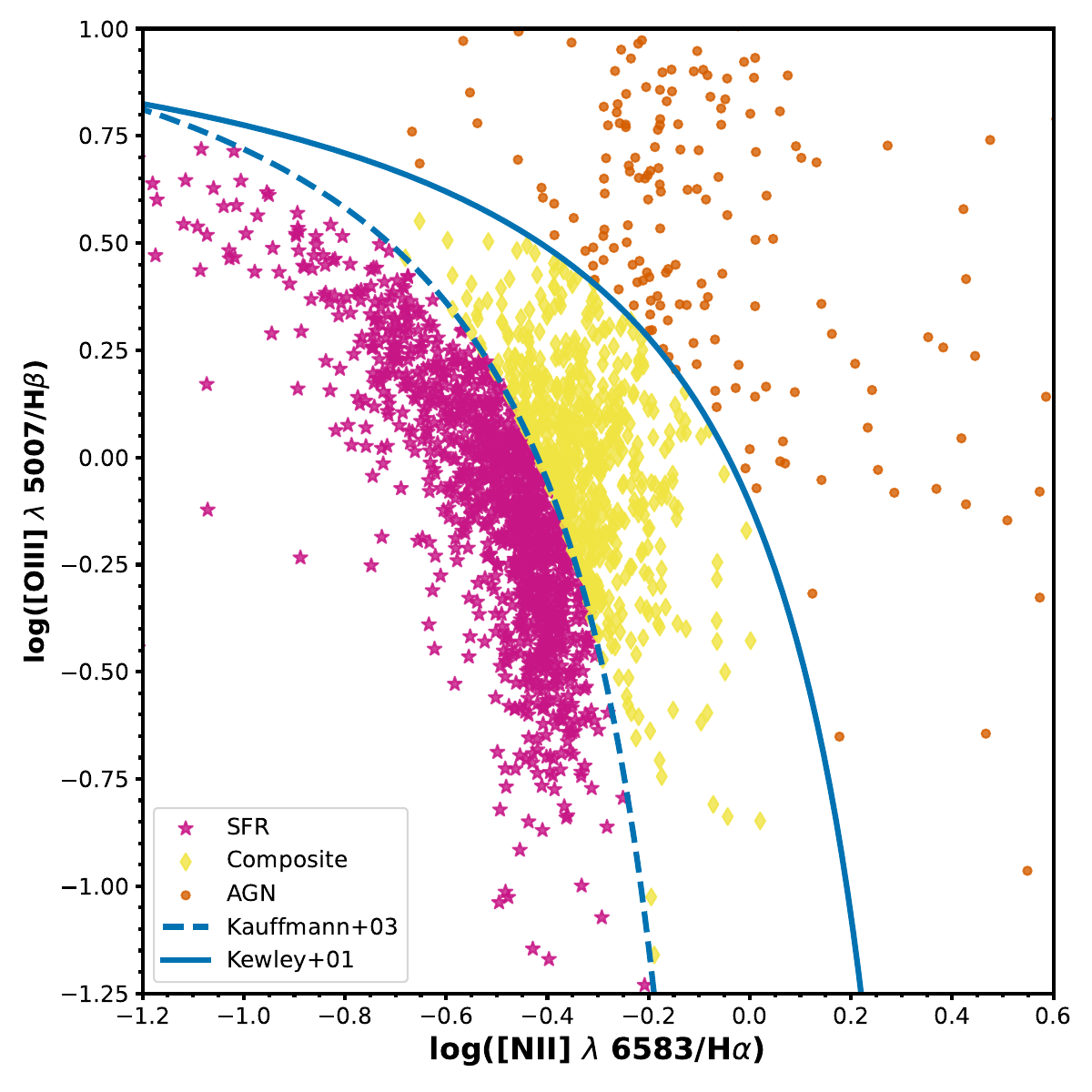}
    \caption{The BPT diagram, which uses the [OIII]/H$\beta$ and [NII]/H$\alpha$ emission line ratios, classifies galaxies as star-forming galaxies (SFGs), active galactic nuclei (AGNs), or composite sources. SFGs, represented by pink stars, lie below the dashed blue Kauffmann line; AGNs, represented by orange circles, are positioned above the solid blue Kewley line; and composite sources, shown as yellow diamonds, are located between the two diagnostic lines. These sources are drawn from the EMU and GAMA catalogues and were processed as described in \S\,\ref{sec:Data}.}
    \label{fig:BPT}
\end{figure}

Finally, the classification of SFGs is done using the standard diagnostic diagram of Baldwin, Phillips, and Terlevich \citep[BPT,][]{1981PASP...93....5B, 1987ApJS...63..295V}. As shown in Figure \ref{fig:BPT},  SFGs are defined as those that fall below the line derived by \cite{10.1111/j.1365-2966.2003.07154.x}, forming the basis of our primary sample. Non-star-forming galaxies (nSFGs), identified as those above the theoretical line proposed by \citet{kewley_theoretical_2001}, are thought to predominantly contain AGN-dominated sources. It is important to note that this classification is not infallible and some AGN sources may still be present within our sample \citep[e.g.,][]{Prathap_al._2024}. The region between \citet{kewley_theoretical_2001} and \citet{kauffmann_host_2003} diagnostic lines are composite sources and removed from the sample in addition to the AGN sources to further mitigate possible contamination. It is noted in \citet{sanchez_beyond_2025} that BPT-based classifications may overestimate the number of SFGs by up to $\sim10\%$. While this level of contamination is unlikely to affect the overall results presented here, it is taken into account when drawing our conclusions. This diagnostic framework enhances the reliability of the star formation rate calculations by focusing on regions of active star formation.

The selection of our final sample is summarised under the following criteria:
\begin{enumerate}
	\item Positive detection of the H$\alpha$ and H$\beta$ Balmer lines. (3931 sources)
\item S/N threshold on the H$\alpha$ and H$\beta$ Balmer lines, S/N(H$\alpha$)>3 S/N(H$\beta$)>3. (1736 sources)
\item Redshift quality is checked for nQ $\geq$ 3. (1736 sources)
\item Positive detection of the H$\alpha$ and H$\beta$ equivalent widths (EW). (1734 sources)
\item Ceiling values for H$\alpha$ and H$\beta$ Balmer lines are removed. (1726 sources)
\item ‘IS\_BEST’ flag is true. (1687 sources)
\item SFGs are classified and extracted using the BPT diagram (AGN and composite galaxies are removed from the sample). (1036 sources)
\end{enumerate} 

This results in a final sample of 1036 SFGs with radio detections in G23.



\subsection{Luminosity Density Compilation}
\label{Sec:lums}
We compiled published H$\alpha$ luminosities from a broad range of studies \citep{tresse_h_1998, shioya_h_2008, sullivan_ultraviolet-selected_2000, morioka_star-forming_2008, fujita_h_2003, stroe_large_2015, pascual_properties_2005, ly_luminosity_2007, gallego_current_1995, sobral_large_2013, hippelein_star_2003, drake_evolution_2013, gomez-guijarro_properties_2016, yan_h_1999, hopkins_star_2000, villar_h-based_2008, moorwood_h_2000, hayes_h-alpha_2010, lee_comparison_2009, sun_first_2023, bollo_h_2023, covelo-paz_h_2024, guo_search_2024}, spanning approximately 30 years of research and covering a wide range of redshifts ($z = 0 - 8$). The inclusion of high-redshift data from JWST enables a test of our novel dust correction method for H$\alpha$ LFs at the highest observed redshifts.

Where possible, we used observed H$\alpha$ LFs uncorrected for dust obscuration, using published tabulated LFs or binned luminosity density values directly. For datasets published only as figures, we digitised the luminosity function and extracted luminosity density values manually. In cases where only dust-corrected values were available, we reversed the applied corrections detailed in the published work. The most commonly used correction was the scaling factor from \cite{calzetti_dust_2000}, which we removed to approximate the uncorrected luminosity. In some cases only the average Calzetti correction value was reported, for which we had to assume this in the de-correction across all samples. If insufficient information was available to de-correct dust corrections, the published data was not included in this work. We also noted the IMF assumption used in each study. For any work that included a SFRD calculation adopting the \citet{salpeter_luminosity_1955} IMF, we applied the conversion factor (0.63) to bring all estimates to a common \citet{chabrier_galactic_2003} IMF following the methodology outlined in \citet{madau_cosmic_2014}. It is also worth noting that all published luminosity functions were fitted with single Schechter functions, and all fit parameters were homogenised to consistent units.

\section{Dust Obscuration Correction}
\label{Dust correction}
The aim of this work is to use the relationship between obscured (H$\alpha$) and unobscured (radio) SFRs in order to find a self consistent method for dust correction where spectroscopy or colour correction methods are not applicable. This need comes from the increasing numbers of large rest-frame UV-optical surveys, where dust corrections are difficult to make. Such is the case when the Balmer emission lines are redshifted too far outside the observable wavelength range or have poor S/N \citep[e.g.,][]{reddy_paschen-line_2023, matharu_first_2023, shapley_jwstnirspec_2023}. 
This section details the process we follow to establish a relationship in the SFR between H$\alpha$ and 1.4~GHz radio wavelength tracers. This relationship enables us to convert uncorrected SFRs into dust-corrected values and thereby recover the corresponding inferred intrinsic H$\alpha$ luminosities. In addition, we explore three alternative models in which the slope of the relationship is varied, providing a means to test its use as a proxy for dust content.

While both H$\alpha$ and radio continuum emission are widely used to trace star formation, they probe different physical processes and timescales. H$\alpha$ emission traces ionising photons from massive stars with lifespans $\lesssim$10 Myr, providing a snapshot of very recent star formation. In contrast, radio emission which is primarily non-thermal synchrotron radiation from supernova remnants, traces slightly earlier populations of massive stars over longer timescales \citep[$\sim$10–100~Myr;][]{kennicutt_star_2012}. The timescale offset means the two tracers may diverge in galaxies with bursty or rapidly changing star formation histories, where the current star formation activity is not well captured by time-averaged radio emission. For example, in recently quenched systems, radio emission can persist for $\sim$100 Myr, leading to an overestimate of the present-day SFR compared to H$\alpha$ \citep{arango-toro_probing_2023}.

In galaxies with relatively continuous star formation over these timescales, typical of many star-forming galaxies in the local universe, the tracers are found to correlate well. \citet{duncan_mosdef_2020} find that the relationship between radio and H$\alpha$ SFRs shows no significant evolution out to redshift $z \sim 2.6$, providing empirical support for their comparability across a substantial fraction of cosmic history. \citet{cook_devilsmighteegamadingo_2024} demonstrate that calibrating radio-based SFRs requires accounting for the intrinsic timescale sensitivity of each tracer, especially when comparing to IR or UV-optical measures.  The full impact of timescale differences across galaxy populations and redshifts remains an open question, and continued observational and theoretical work will be key to understanding where and when these tracers may diverge.
The choice of SFR calibration introduces an additional source of uncertainty, as it has been shown to significantly impact both SFR and SFRD estimates, particularly at higher luminosities \citep[e.g.,][]{matthews_cosmic_2021}. In this work we adopt the calibrations of \citet{kennicutt_star_1998} and \citet{bell_estimating_2003}, though alternative empirical relations could also be explored. With sufficiently broad photometric coverage, spectral energy distribution fitting provides another avenue for deriving consistent SFR estimates.

Despite these limitations, a SFR-dependent obscuration correction offers promising potential as a flexible and observationally grounded tool for estimating star formation rates, especially in regimes where traditional dust corrections are inaccessible or unreliable. Unlike fixed attenuation laws or colour-based proxies, this method directly links two widely observable quantities: observed H$\alpha$ luminosity, which is sensitive primarily to the unobscured component of recent star formation (and requires extinction corrections to recover the total), and 1.4\,GHz radio luminosity, which is largely insensitive to dust and therefore traces the total star formation (both obscured and unobscured). When calibrated properly, this approach could bridge datasets from UV-optical and radio surveys in a self-consistent way and be deployed in large-scale survey pipelines where full SED fitting is not feasible.

\subsection{Radio Star Formation Rates}
A common approach in computing radio-based SFRs is to use the flux density at $1.4$\,GHz. Since the early science EMU catalogue provides data at $888$\,MHz, we convert to $1.4$\,GHz assuming a spectral index\footnote{We adopt the convention $S\propto \nu^{\alpha}$.} of $\alpha=-0.7$. Varying this assumption has a negligible effect on the final result.
Once these conversions are made, the radio luminosity can be calculated using:
\begin{equation}
    L_{1.4GHz}=\frac{4\pi \times D_L^2 \times S_{1.4}}{({1+z})^{1+\alpha}},
\end{equation}
where $D_L$ [pc] is the luminosity distance, $S_{1.4}$ [\textrm{W~Hz}$^{-1}$] is the radio flux, and $z$ is the redshift.

The radio luminosity is first applied in the calibration derived by \citet{2003ApJ...586..794B}, which is based on the correlation between radio and FIR emission \citep{2003ApJ...586..794B, 2003ApJ...599..971H}, $\text{SFR}_{\textrm{1.4~GHz}}[\mathrm{M}_{\odot}\,\mathrm{yr}^{-1}]=$

\begin{equation}
\begin{cases}
\dfrac{L_{\textrm{1.4~GHz}}~[\textrm{W}~\textrm{Hz}^{-1}]}{1.81 \times 10^{21}} & \text{for } L > L_c, \\
\dfrac{L_{\textrm{1.4~GHz}}~[\textrm{W}~\textrm{Hz}^{-1}]}{[0.1 + 0.9(L/L_c)^{0.3}] \cdot 1.81 \times 10^{21}} & \text{for } L \leq L_c,
\end{cases}
\label{bell}
\end{equation}
where $L_c=6.4\times 10^{21} ~\mathrm{W~Hz}^{-1}$. Bell uses a step-function-based correction based on a galaxy’s 1.4\,GHz luminosity. The luminosity traces non-thermal synchrotron emission from supernova remnants, which correlates with recent star formation. The correlation is strong in star-forming galaxies but less reliable in quiescent galaxies, where the emission may originate from old stars or AGN. The step function in Eq.~\ref{bell}, helps avoid overestimating the star formation rate in such cases. It is worth noting that the \citet{bell_estimating_2003} method remains broadly consistent other radio-SFR based calibrations, typically within ~20–30\%, depending on assumptions such as IMF, radio frequency, and treatment of thermal emission \citep[e.g.,][]{murphy_calibrating_2011, delhaize_vla-cosmos_2017}. As radio wavelengths are unobscured by dust, this SFR acts as the naturally dust corrected SFR. 

\subsection{H$\alpha$ Star Formation Rates}
The SFRs from H$\alpha$ emission are derived in order to compare with their radio counterpart. Eq.~\ref{LumEq} and \ref{halpsfr} are first applied to correct for several observational systematics. The H$\alpha$ luminosity is then calculated following the methods outlined in \citet{2003ApJ...599..971H} and \citet{2011MNRAS.415.1647G}. Including correcting for stellar absorption and for the aperture losses associated with fibre spectroscopy. We do not apply a dust extinction correction here in order to replicate conditions where such steps are unavailable, thus Balmer Decrement corrections are omitted \citep{gunawardhana_galaxy_2013}. 
\begin{multline}
\label{LumEq}
     L_{H\alpha}~[\textrm{W}~\textrm{Hz}^{-1}]=(EW_{H\alpha}+EW_c) \times 10^{-0.4(M_r -34.10)} \\ \times \frac{3 \times 10^{11}}{[6564.61(1+z)]^2}.
\end{multline}

The common equivalent width correction (EW$_c$) is assumed to be 1.3$\, \mathring{\text{A}}$, \citep{gunawardhana_galaxy_2013}. The absolute magnitude in the $r$ band (M$_r$) is provided by the GAMA survey.  

The H$\alpha$ luminosity is then used to estimate uncorrected SFR as derived by \citet{1998ARA&A..36..189K}: 

\begin{equation}
\label{halpsfr}
    \text{SFR}_{H\alpha}[M_{\odot}\,yr^{-1}]= \frac{L_{H\alpha}~[\textrm{W}~\textrm{Hz}^{-1}]}{7.9 \times 10^{35}}.
\end{equation}

\subsection{SFR Relationship}
\begin{figure}
    \centering

    \includegraphics[width=\linewidth,trim=20 25 5 5]{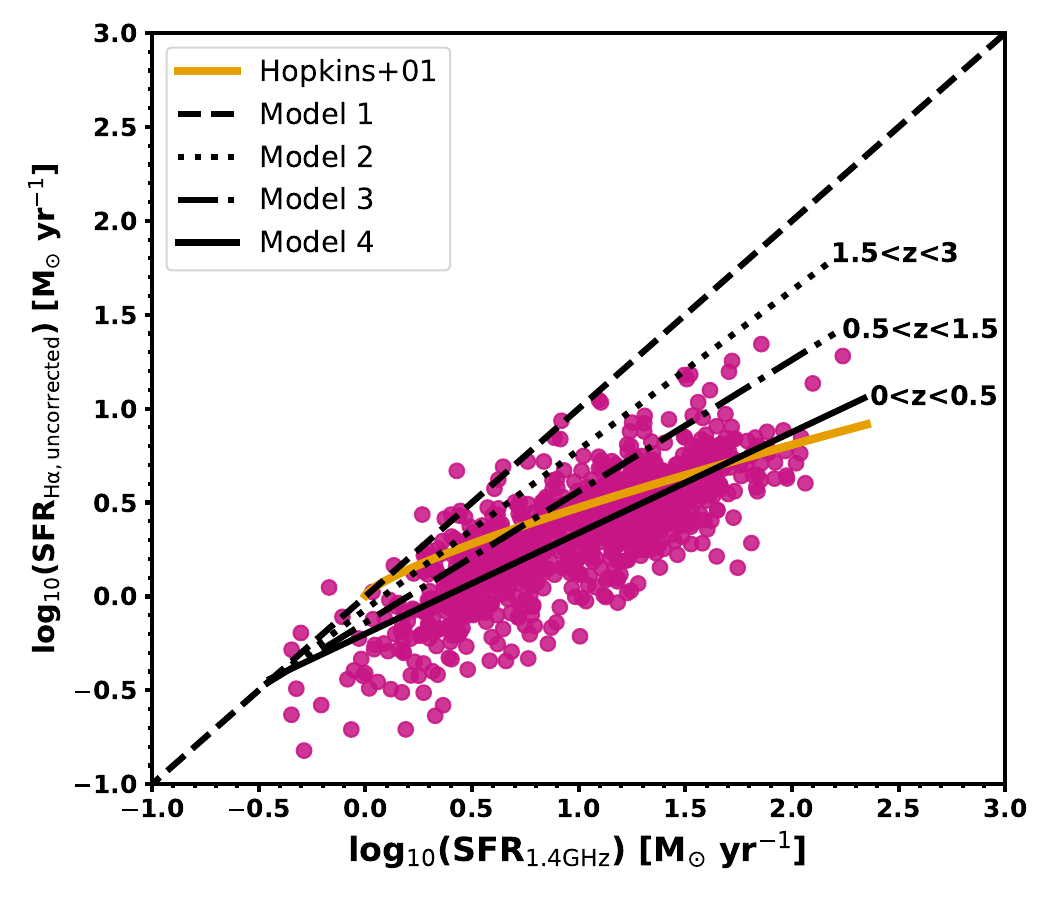}
    \caption{The relationship between H$\alpha$ and $1.4$\,GHz radio tracers of star formation. Four models are shown: the solid line is the best-fit relation to the local EMU-GAMA SFG sample, while the dashed line marks the 1:1 case, where H$\alpha$ and radio SFRs would be equal, making the 1:1 line the dust free line. The dotted and dot-dashed lines represent proposed interim models. For each model, the redshift interval over which its dust-corrected results align with the published SFRD measurements from Figure~\ref{fig:CSFD} is indicated. The truncated equation for each model is shown here for reference but can be seen in full in Table~\ref{Tab: Models}. The calibration from \citet{hopkins_toward_2001} is shown in orange.}
    \label{fig:SFRComp}
\end{figure}

Where optical dust correction methods are unavailable or unreliable, the relationship between uncorrected H$\alpha$ SFR and radio SFR is explored as a means for dust obscuration correction.
Once the SFRs for each of our SFGs are calculated, they are compared in Figure~\ref{fig:SFRComp}. The \citet{bell_estimating_2003} and \cite{kennicutt_star_1998} SFR calibrations are adopted here but the calibration choice has minor effects within the overall scatter. 
To determine the relationship between the SFR estimates, we performed a linear regression using \texttt{numpy.polyfit} with cov=True, which minimises the squared error and returns the covariance matrix for estimating parameter uncertainties. For low SFRs, the 1:1 relationship is adopted as in \citet{hopkins_toward_2001}, and fitted relationship only begins where data becomes available. The best-fit relation between H$\alpha$ and radio SFRs is shown as the solid line in Figure~\ref{fig:SFRComp}, with the fitted equation given as:
\begin{equation}
\label{FittedSFR}
\log_{10}(\mathrm{SFR}_{\mathrm{H}\alpha,\mathrm{uncorrected}}) = 0.55\log_{10}(\mathrm{SFR}_{1.4\mathrm{GHz}}) - 0.22.
\end{equation}
The standard errors on the slope and intercept are both $\pm$0.02, as estimated from the diagonal of the covariance matrix.

Figure~\ref{fig:SFRComp} is constructed using data from the local Universe, $0.0 < z < 0.35$, $0.01~\mathrm{M}_\odot~\mathrm{dex}^{-1} <M < 11.69~\mathrm{M}_\odot~\mathrm{dex}^{-1}$, and the fit begins at the onset of available data, following the approach of \cite{hopkins_toward_2001}. To evaluate the applicability of this method across evolving redshift, three additional SFR relationship models are developed. The first, Model 1 (shown as dashed line in Figure~\ref{fig:SFRComp}), assumes no correction—treating the SFRs derived from H$\alpha$ and $1.4$\,GHz radio as equivalent. The other two models have slopes that lie between the 1:1 case and our fitted dust-correction model. These models are summarised in Table~\ref{Tab: Models} where the fitted relationship shown in Equation~\ref{FittedSFR} is now referred to as Model 4. 

The physical interpretation of these models relates to the inferred level of dust attenuation. The 1:1 line (Model~1) serves as a proxy for a dust-free Universe. In contrast, the fitted relationship (Model~4) reflects the dust content of the local Universe. Models~2 and 3, which fall between Models~1 and 4, therefore represent scenarios with intermediate—i.e., reduced—dust attenuation. Our correction method works by inferring the level of dust
attenuation from these models and correcting for it accordingly.

\begin{table}[!ht]
    \centering
    \begin{tabular}{cc} \hline \hline
        \textbf{Model} & \textbf{Relationship} \\ \hline
        1 & log(SFR$_{\mathrm{H}\alpha,\mathrm{uncorrected}}$) = log(SFR$_{1.4\mathrm{GHz}}$) \\ 
        2 & log(SFR$_{\mathrm{H}\alpha,\mathrm{uncorrected}}$) = $0.85$\,log(SFR$_{1.4\mathrm{GHz}}$) $-\, 0.07$ \\ 
        3 & log(SFR$_{\mathrm{H}\alpha,\mathrm{uncorrected}})$ = $0.7$\,log(SFR$_{1.4\mathrm{GHz}}$) $-\, 0.14$  \\ 
        4 & log(SFR$_{\mathrm{H}\alpha,\mathrm{uncorrected}}$) = $0.55$\,log(SFR$_{1.4\mathrm{GHz}}$) $-\, 0.22$ \\ \hline
    \end{tabular}
    \caption{Fitted models for the relationship between H$\alpha$ and 1.4GHz radio wavelength tracers. Here Model 1 is the 1:1 case with no correction, Model 4 is the fitted relationship to the data and Models 2 and 3 are two intermediate cases that have been chosen to fit SFRD data presented in \S\ref{Results}. }
    \label{Tab: Models}
\end{table}

To apply this relationship for dust correction, we first convert the H$\alpha$ luminosities that are uncorrected for dust into uncorrected SFRs using the calibration from \citet{kennicutt_star_1998}, adjusted with the Chabrier IMF correction from \citet{madau_cosmic_2014}. We then determine the corresponding dust-corrected SFR by finding its corresponding radio SFR, dust corrected, based on the model slope applied. The corrected SFR is subsequently converted back into a luminosity by reversing the same SFR calibration. The resulting H$\alpha$ luminosities are used to fit new luminosity functions, as detailed in \S\ref{Results}.

An alternative way to express our luminosity- and redshift-dependent calibration from uncorrected H$\alpha$ SFRs to their radio-derived counterparts is through a multiplicative correction factor (CF). This factor is defined, for each of the three redshift ranges over which our models are applicable, as the ratio of the radio SFR to the uncorrected H$\alpha$ SFR. Correction factors are computed for four representative uncorrected H$\alpha$ luminosities and are presented in Table~\ref{tab:ha_correction_factors}. This quantity effectively parameterises the deviation of each calibration from the one-to-one relation between H$\alpha$ and radio SFRs. The resulting correction factors clearly indicate that progressively smaller dust corrections are required at higher redshifts.

\begin{table}[ht]
\centering
\caption{Correction factors converting uncorrected H$\alpha$ star formation rates to radio-derived star formation rates as a function of redshift and uncorrected H$\alpha$ luminosity. Correction factors are defined as the ratio $\mathrm{SFR}_{\mathrm{1.4GHz}} / \mathrm{SFR}_{\mathrm{H}\alpha, \textrm{uncorrected}}$ and are shown for four representative uncorrected H$\alpha$ luminosities. Luminosities are given in units of $10^{36}\,\mathrm{WHz^{-1}}$.}
\label{tab:ha_correction_factors}

\begin{tabularx}{0.8\linewidth}{lXXXX}
\toprule
\toprule
\textbf{Redshift} &
\multicolumn{4}{c}{\textbf{Uncorrected H$\alpha$ Luminosity ($10^{36}\,\mathrm{W\,Hz^{-1}}$)}} \\
\cmidrule(lr){2-5}
 & \textbf{0.25} & \textbf{0.8} & \textbf{2.5} & \textbf{8.0} \\
\midrule
0--0.5   & 1.0 & 2.5 & 6.5 & 16.7 \\
0.5--1.5 & 1.0 & 1.6 & 2.6 & 4.3  \\
1.5--3.0 & 1.0 & 1.2 & 1.5 & 1.8  \\
\bottomrule
\end{tabularx}

\end{table}

\section{Results and Analysis}
\label{Results}
Once we applied our new SFR-based dust correction model to the published luminosities that were either originally uncorrected, or first de-corrected to remove prior extinction estimates, we fit new LFs to the data. Allowing us to compare our approach with existing results and to evaluate how the relationship may vary with redshift by integrating to obtain the SFRD. Using the dust correction from the fitted relationship (Model 4) we present updated LFs in Figure~\ref{LFs}. The luminosity values when using the radio-optical SFR based relationship as a dust correction are substantially brighter than in the original LFs, sitting up to two orders of magnitudes higher at all redshifts, compared to the original data. 

\subsection{Fitting new luminosity functions}
The LF, $\Phi(L)$, describes the number of galaxies per unit volume per unit luminosity interval. It is a fundamental statistical tool in observational cosmology that characterises the distribution of galaxy luminosities across the Universe. Since galaxy luminosity is correlated with star formation activity, the LF is often used to study how star formation varies with galaxy population and redshift.

A widely used analytical form for modelling galaxy LFs in the optical regime is the Schechter function, introduced by \citet{schechter_analytic_1976}, and given by:
\begin{equation}
\phi(L)\, dL = \phi^* \left( \frac{L}{L^*} \right)^{\alpha} 
\exp\!\left(-\frac{L}{L^*}\right) \frac{dL}{L^*},
\end{equation}
where $\phi^*$ is the normalisation factor for the number density of galaxies, $L_*$ is the characteristic luminosity where the function transitions from a power-law to an exponential cutoff, and $\alpha$ is the faint-end slope, describing the abundance of low-luminosity galaxies. We adopt the Schechter function approach to fit the newly corrected luminosity functions. 

\begin{figure*}[t]
    \includegraphics[width=\linewidth,trim=20 25 5 5]{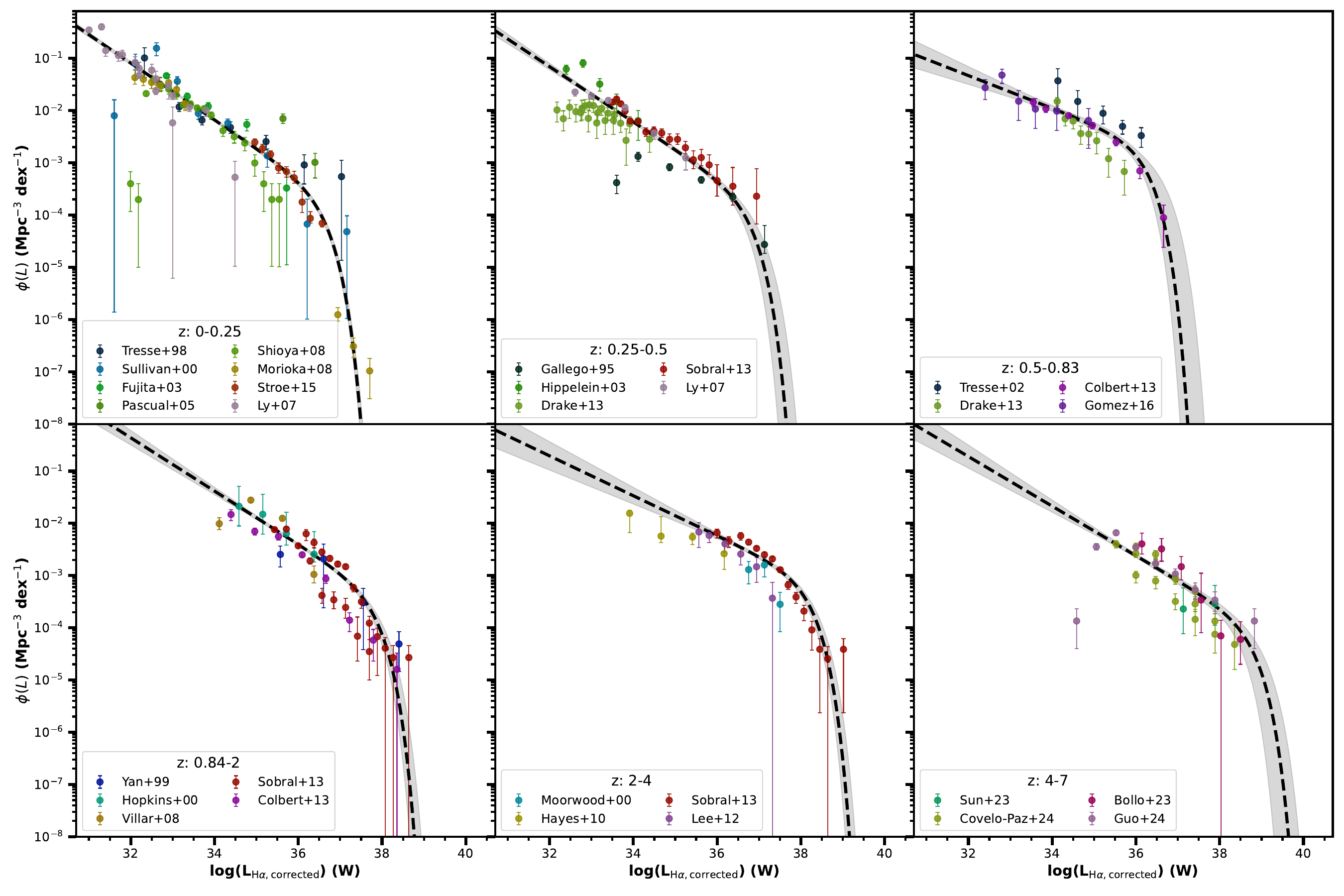}
    \caption{Published luminosity values, corrected for dust obscuration using the fitted SFR relation from Figure~\ref{fig:SFRComp} (coloured symbols), along with the corresponding best-fit Schechter functions (dashed line), shown across six redshift bins. Where possible, raw luminosity data from these sources have been used; if necessary, published dust corrections were removed. Our fitted SFR-based dust correction relation (Model 4) was applied instead and new LFs were refitted, where the shaded region depicts the 95\% confidence interval from 100,000 MCMC iterations. See \S\ref{Sec:lums} for relevant published H$\alpha$ luminosity data. }
    \label{LFs}
\end{figure*}

To fit Schechter luminosity functions, the luminosity data were divided into six redshift bins spanning $z = 0$\,--\,8. These bins were chosen to group together data with similar redshifts while maintaining a sufficient density of points within each bin to enable statistically robust fitting. The redshift bins are shown in Figure~\ref{LFs}, where a reasonably consistent trend is observed within each bin. Although some data points appear to be outliers, they are retained in the fitting process.
While the luminosity function refitting process accounts for reported uncertainties, other factors—such as survey area—can influence the results. Larger survey volumes increase the likelihood of detecting rarer, high-luminosity sources, which in turn affects the fitted values of parameters. This becomes especially significant in redshift bins where observational scatter is high.

Each redshift bin was initially fitted with a Schechter function by minimising the residuals, weighted by the uncertainties of the data points. To more reliably quantify the uncertainties in the fitted parameters, the initial solution was refined using a Markov Chain Monte Carlo (MCMC) approach implemented with the Python package \texttt{emcee}, which uses MCMC to efficiently explore parameter space and estimate uncertainties for complex models \citep{foreman-mackey_emcee_2012}.\footnote{\url{https://emcee.readthedocs.io/}}
 The MCMC was run for 100,000 iterations for each redshift bin. The fitting procedure was repeated independently for the luminosity data corresponding to each of the four models. The best-fit luminosity function parameters for the local SFG SFR dust correction relation (Model 4) are reported in Table~\ref{LF params}.

\begin{table}[]
\begin{tabular}{cllll}
\hline
\hline
\textbf{\begin{tabular}[c]{@{}c@{}}Redshift\\ $z$\end{tabular}} &
\multicolumn{1}{c}{\textbf{\begin{tabular}[c]{@{}c@{}}Log $\phi^*$\\ (Mpc$^{-3}$)\end{tabular}}} &
\multicolumn{1}{c}{\textbf{\begin{tabular}[c]{@{}c@{}}Log $L^*$\\ (W)\end{tabular}}} &
\multicolumn{1}{c}{\textbf{$\alpha$}} &
\multicolumn{1}{c}{\textbf{$\chi^2$}} \\ \hline
$0.0-0.25$  & $-3.92 \pm 0.03$ & $36.54 \pm 0.02$ & $-1.54 \pm 0.01$ & $1.44$ \\
$0.25-0.5$  & $-4.01 \pm 0.09$ & $36.71 \pm 0.12$ & $-1.53 \pm 0.02$ & $0.79$ \\
$0.5-0.83$  & $-2.99 \pm 0.10$ & $36.17 \pm 0.13$ & $-1.31 \pm 0.04$ & $0.36$ \\
$0.83-2.0$  & $-3.70 \pm 0.06$ & $37.79 \pm 0.07$ & $-1.51 \pm 0.02$ & $1.00$ \\
$2.0-4.0$   & $-3.41 \pm 0.06$ & $38.14 \pm 0.06$  & $-1.38 \pm 0.03$ & $0.74$ \\
$4.0-7.0$   & $-4.27 \pm 0.12$ & $38.72 \pm 0.17$ & $-1.47 \pm 0.03$ & $0.92$  \\ \hline
\end{tabular}
\caption{Best-fit parameters of H$\alpha$ Luminosity function for maximum dust correction model, including reduced chi-squared ($\chi^2$) values for each redshift bin.}
\label{LF params}
\end{table}

\subsection{Calculating cosmic star formation rate density}
We can derive the cosmic SFRD, $\rho_{\text{SFR}}$, via integration of the luminosity function to compute the luminosity density, $\rho_L$:
\begin{equation}
\rho_{\text{L}} = \int_{L_{\mathrm{lim}}}^{\infty} L \Phi(L) dL = \phi^* L^* \Gamma(\alpha + 2, L_{\mathrm{lim}} / L^*),
\end{equation}
where $\Gamma$ is the incomplete gamma function, and $L_{\text{lim}}$ is the lower luminosity limit of the integration which is often set by survey completeness \citep{madau_cosmic_2014}. We adopt a value of $L_{\text{lim}}=10^{30}~\mathrm{W}$ to account for the full range of the data. Through testing of varying lower luminosity limits, there was minimal change in final SFRD value, with up to $\approx$~2\% difference when using values up to $L_{\text{lim}}=10^{36.6}~\mathrm{W}$ and no additional value when integrating to $L_{\text{lim}}=10^0~\mathrm{W}$. 

The luminosity density can then be converted into an estimate of the SFRD using a wavelength-dependent SFR calibration factor. These calibrations depend on the IMF and the choice of SFR tracer. For the H$\alpha$ luminosity we adopt the\citet{kennicutt_star_1998} calibration, converted to an assumption of a Chabrier IMF:
\begin{equation}
\rho_{\text{SFR}}~  (\mathrm{M}_\odot~ \mathrm{yr}^{-1}\mathrm{Mpc}^{-3}) = 7.9 \times 10^{-35.24} 
~\rho_{\text{L}}~(\mathrm{W}~\text{Mpc}^{-3}).
\end{equation}

\begin{figure*}[t]
    \centering
    \includegraphics[width=0.98\linewidth,trim=20 20 5 5]{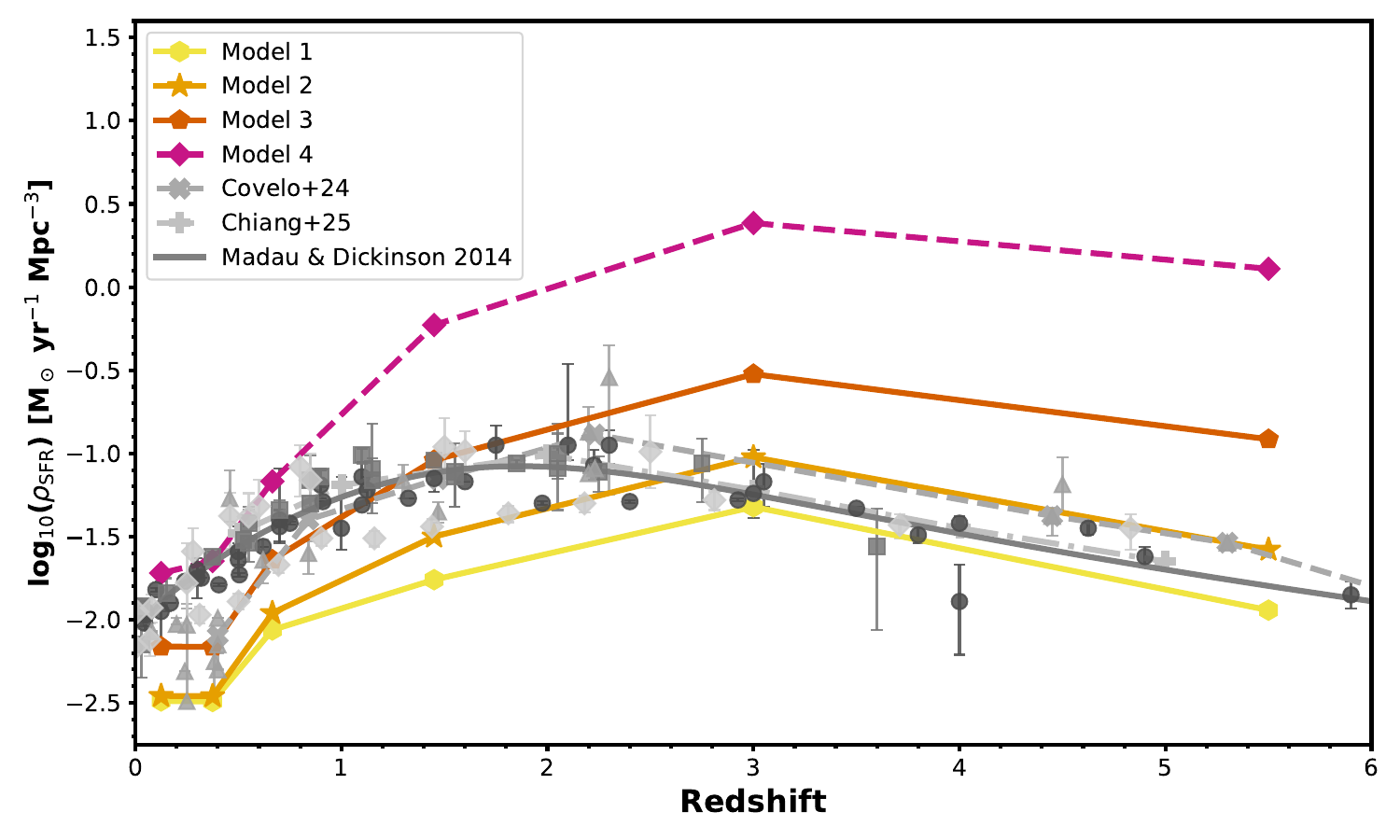}
    \caption{Cosmic SFRDs derived from dust-corrected luminosity functions using each of our four dust correction models, shown in Table~\ref{Tab: Models}. These results are compared with recent dust-corrected measurements from \citet{covelo-paz_h_2024} and \citet{chiang2025cosmicinfraredbackgroundtomography} in dashed and dot-dashed linesrespectively. The \citet{madau_cosmic_2014} relation is also shown in solid grey for comparison. Published dust-corrected star formation rate density (SFRD) values are shown in grey, with markers indicating the observational tracer: UV (circles), H$\alpha$ (triangles), infrared (squares), and radio (diamonds). The published data is compiled in Table \ref{Tab:SFRD_combined}.}
    \label{fig:CSFD}
\end{figure*}

The cosmic SFRD provides a valuable means of tracing the evolution of SFR over cosmic time. The SFRD evolution for each of our four models is presented in Figure~\ref{fig:CSFD}. This figure compares our results against single or compiled dust-corrected SFRDs from 33 published sources across UV, H$\alpha$, IR and Radio wavelengths, see Table \ref{Tab:SFRD_combined}. As well as the recent dust corrected measurements from \citet{covelo-paz_h_2024} and \citet{chiang2025cosmicinfraredbackgroundtomography}. 

Figure~\ref{fig:CSFD} demonstrates that the fitted relationship between SFR tracers aligns well with dust-corrected observational data at low redshifts, which is expected given that the relationship was derived from local-Universe data. However, at higher redshifts ($z > 1$), the SFRD evolution predicted by the fitted relation exhibits an extreme overestimate that is clearly not well supported by existing measurements.

The less extreme models of Figure~\ref{fig:SFRComp} demonstrate progressively reduced overestimates at higher redshifts, but also underestimates at the lowest redshifts. The model where no dust correction is applied is the most underestimated but at redshifts $z > 3.5$ there is H$\alpha$ data that is well represented by a dustless model. This suggests that an approach to dust correction that evolves with redshift is likely to be necessary over differing redshift ranges.

\begin{figure*}
    \centering
    \includegraphics[width=\linewidth,trim=20 20 5 5]{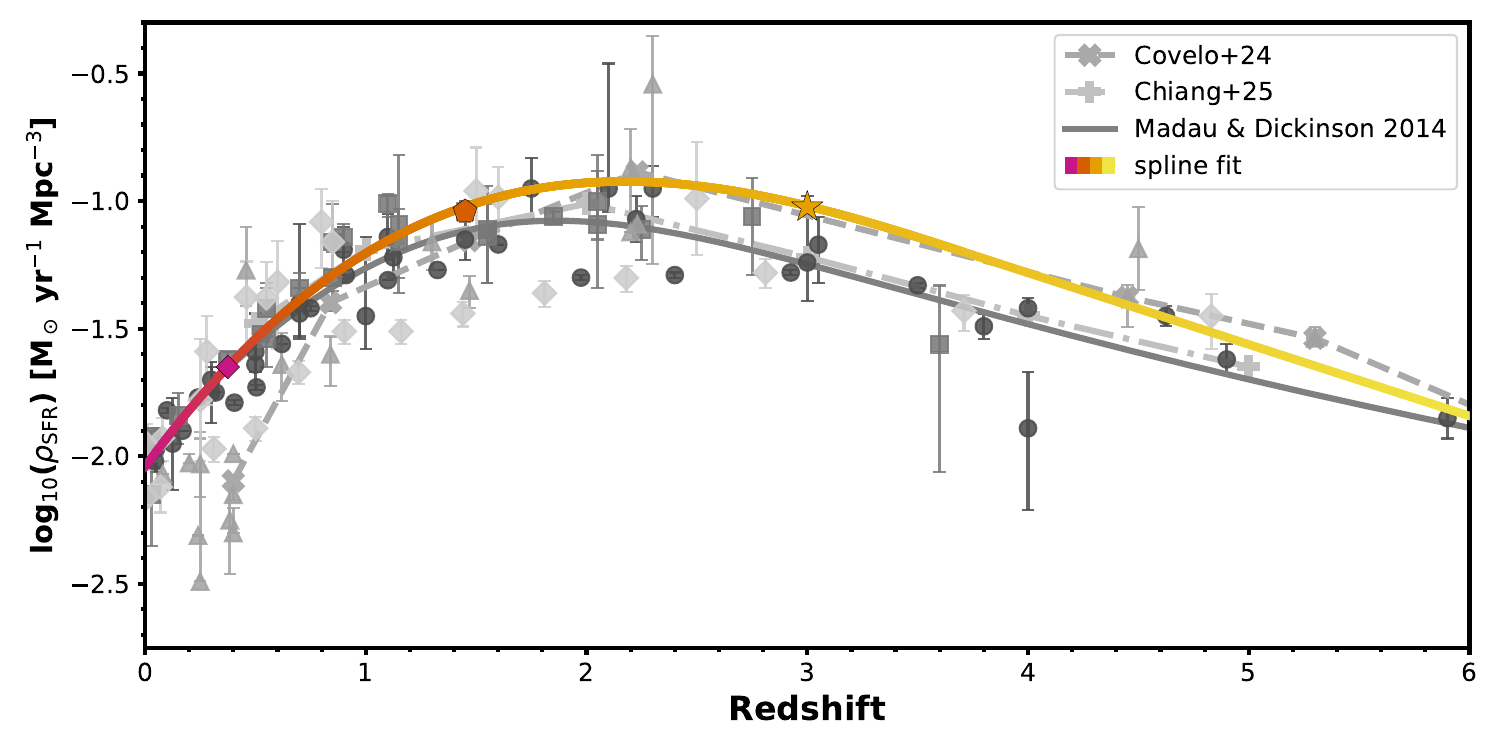}
    \caption{The SFRD from a spline fit to the step-function model for the SFRD evolution, illustrating the evolving dust correction derived from the fitted relationship between H$\alpha$ and radio SFR tracers. Here the symbols are the same as in Figure \ref{fig:CSFD}. The colors of the spline fit correspond to those in Figure \ref{fig:CSFD} for the model proposed in each redshift range, from the 'local Universe dust content' scenario of Model 4 to the 'dust-free' scenario of Model 1. The data points where the spline fit changes model are shown by the overlayed markers, where the marker type represents the model as shown in Figure\ref{fig:CSFD}.}
    \label{fig:CSFD_trend}
\end{figure*}

For redshifts up to $z \approx 0.5$, the SFR-dependent obscuration correction, calibrated using EMU and GAMA observations over $0.0 < z < 0.35$, produces a cosmic SFRD that is consistent with published values across this range. This results demonstrates that the method can serve as a promising alternative for estimating SFRs in the local Universe, particularly in cases where traditional obscuration corrections based on spectral line diagnostics or galaxy colour methods are unavailable. Such conditions are common in large photometric or radio surveys where emission line data may be incomplete or of low quality \citep{prathap_emugama_2025}.

Beyond $z \sim 1$, the SFR-dependent obscuration correction increasingly overestimates the SFRD, (Figure~\ref{fig:CSFD}), with deviations reaching up to two orders of magnitude compared to published work (e.g. \citealt{madau_cosmic_2014}, \citealt{bouwens_uv_2015}). As illustrated in Figure~\ref{fig:CSFD}, suggesting that the linear relationship between obscured H$\alpha$ luminosities and 1.4\,GHz radio luminosities observed locally does not persist out to redshifts $z \sim 7$. 


As the redshift increases toward $z \approx 1.5$, the originally calibrated linear relation begins to overestimate the dust correction. This results in a CSFD that is significantly higher than reported values. In contrast, a steeper slope, such as that used in Model 3 (shown in dark orange in Figure~\ref{fig:CSFD}), produces a markedly better match to the observed data. This suggests that, at progressively increasing redshifts, there is progressively less dust obscuration than seen in the local Universe.


At redshifts approaching $z \approx 3.0$, we enter the epoch of peak cosmic star formation activity, often referred to as “cosmic noon”, \citep{madau_cosmic_2014}. In this regime, we find that the SFR-dependent obscuration correction requires a slope intermediate between that of Model 2 and Model 3 (shown in light and dark orange, respectively, in Figure~\ref{fig:CSFD}) in order to approximate published values for the SFRD. Implying a transitional phase in the relationship between H$\alpha$ and radio luminosities, possibly reflecting a balance between increasing dust production in rapidly evolving galaxies and the onset of more massive, dust-rich systems. 

The redshift range above $z \approx 3.0$, is marked by considerable scatter among different SFR tracers, reflecting a lack of consensus across methodologies \citep{bouwens_lower-luminosity_2012, schenker_uv_2013, novak_vla-cosmos_2017, bollo_h_2023, covelo-paz_h_2024}. The variations may stem from uncertainties in dust content, evolving ISM geometries, dust build up timescales and selection biases inherent in current high-redshift galaxy samples. Recent JWST results also suggest that the emergence of a distinct population of red star-forming galaxies (RedSFGs) at $3<z<4$ may play a role in shaping this behaviour. \citet{tarrasse_compact_2025} identify these RedSFGs as a possible evolutionary link between star-forming and quiescent galaxies, potentially contributing to the observed slope transition in this redshift range.
It is worth noting that the \citet{madau_cosmic_2014} SFRD for $z>3$ aligns almost identically with our Model~1, implying no dust obscuration at all. This is not likely to be physical, given the results noted above, and many others that measure non-negligible obscuration at high redshift.

In addition to observational factors, the evolution of galaxy metallicities with redshift may also play a role \citep{panter_star_2003, camps-farina_chemical_2022}. Lower-metallicity galaxies, which are more common at higher redshifts, produce lower star-formation rates for a given H$\alpha$ luminosity. Therefore by applying SFR–luminosity calibrations derived for predominantly solar-metallicity for these high-redshift systems could introduce a systematic bias. While this effect is expected to be smaller than the observed H$\alpha$ deficit, it may still contribute to the discrepancy; however, quantifying its impact lies beyond the scope of this work.

The fact that neither the steeper nor shallower correction alone accurately reproduces the observed star formation density implies that a single, static obscuration correction is inadequate across this redshift range. Instead, the obscuration characteristics may be evolving more dynamically. This is potentially influenced by variations in star formation modes, ISM geometry \citep{dalcanton_formation_2004, holwerda_evolution_2012, holwerda_frequency_2019}, or dust-to-gas ratios.
This highlights the need for either a smoothly evolving SFR–based dust calibration or a multi-parametric approach that accounts for additional galaxy properties beyond redshift alone.

\subsection{Composite Model}
These results indicate that the SFR-based dust correction must evolve with redshift to match optical based approaches, no single fit is appropriate. A composite model, constructed by combining the most consistent model across different redshift ranges, is proposed to describe the SFRD evolution. The models are used in with spline fit and then smoothed to create a sample model which is shown in Figure~\ref{fig:CSFD_trend}.

Here we adopt Model 4 in the redshift range $0 < z < 0.5$, based on the H$\alpha$ and Radio SFR relationship in our local Universe which uses observational data in this redshift region. Then a smooth transition to Model 3 is applied for $0.5 < z < 1.5$. Then Model 2 out to redshift $z \approx 3$. In this redshift range, the Model should sit somewhere between Model 1 and 2. The redshift ranges over which each model best aligns with the observed cosmic SFRD are annotated in Figure~\ref{fig:SFRComp}. This composite model is expressed as a redshift- and luminosity-dependent correction factor in Equation~\ref{eq:CF}, where $\kappa = 7.9\times10^{35}$ is the SFR calibration constant from \citet{kennicutt_star_1998}.
\begin{equation}
\mathrm{CF}(L_{\mathrm{H}\alpha}, z) =
\begin{cases}
10^{0.4}
\left(\dfrac{L_{\mathrm{H}\alpha}}{\kappa}\right)^{-2.22}, & 0 < z < 0.5, \\[5pt]

10^{0.2}
\left(\dfrac{L_{\mathrm{H}\alpha}}{\kappa}\right)^{-3.33}, & 0.5 \le z < 1, \\[5pt]

10^{0.08}
\left(\dfrac{L_{\mathrm{H}\alpha}}{\kappa}\right)^{-6.67}, & 1 \le z < 3.
\end{cases}
\label{eq:CF}
\end{equation}
In practice, our prescription for applying a luminosity- and redshift-dependent dust obscuration correction is as follows: first, correct the observed H$\alpha$ luminosity using Equation~\ref{LumEq}; next, convert this corrected luminosity into an uncorrected H$\alpha$ SFR via Equation~\ref{halpsfr}; finally, apply the correction factor from Equation~\ref{eq:CF} to obtain a radio-derived SFR that accounts for dust extinction.

The revised model aligns more closely with recent results from \citet{covelo-paz_h_2024} and \citet{chiang2025cosmicinfraredbackgroundtomography}, reinforcing the value of a composite approach. In this work, published dust-corrected luminosity functions at $z>0.35$ are treated as a reference standard. This is not because they represent an absolute ground truth, but because they are derived from well-established, empirically tested correction techniques. Our aim is to demonstrate that the novel radio-based method proposed here could reproduce similarly reliable results. These findings highlight the potential of a radio–optical hybrid framework, and suggest that a comprehensive study of SFR-based dust correction methods over cosmic time could be an important step toward developing a SFR based dust correction technique into a robust and widely trusted complement to traditional optical approaches.
The development of an evolving SFR-dependent obscuration correction could also offer valuable insights into the physical evolution of dust and gas in galaxies over cosmic time. By examining how the relationship between H$\alpha$ and radio emission changes, one can indirectly track variations in dust content \citep{driver_gamag10-cosmos3d-hst_2018}. This model contrasts with the observation that the average dust mass at redshift $z \approx 1$ is approximately a factor of $\sim 3$ higher than at $z \approx 0$. One possible explanation lies in how the spatial distribution of dust within galaxies may change with increasing redshift \citep{farrah_nature_2008, pope_mid-infrared_2008, casey_are_2014, farley_galaxy_2025}. If regions with higher dust mass do not spatially coincide with areas of strong H$\alpha$ emission, then global dust correction methods may introduce biases, leading to an over- or underestimation of the true obscured star formation. The work of \citet{farley_galaxy_2025} also investigates the scatter of BD among the GAMA sample and how dust mass can further constrain dust quantity in SFGs. Part of the discrepancy may also arise from an increasing contribution of low-luminosity AGN to the radio or FIR emission in galaxies at $z \gg 1$, which is explored in \citet{thomson_evolution_2017}.

\section{Conclusion}
\label{sec:conclusion}

In this work, we have explored the viability of a SFR-dependent obscuration correction based on the empirical relationship between the obscured H$\alpha$ SFR tracer and the unobscured 1.4~GHz radio SFR tracer. The local SFG H$\alpha$-radio SFR relationship was initially calibrated as a linear fit over the low-redshift range $0.0 < z < 0.35$, using EMU and GAMA data, and subsequently applied as a correction to H$\alpha$ luminosities across a broad compilation of published luminosity values. Once our new dust correction method was applied, the new LFs were refitted with Schechter functions. The dust-corrected corrected LFs were then used to estimate the SFRD, $\rho_{\text{SFR}}$, which was compared to other published trends.

At low redshift, the SFR-based dust correction reproduces the cosmic SFR density in good agreement with previous studies. However, at $z>1$ it significantly overestimates $\rho_{\mathrm{SFR}}$, by up to two orders of magnitude. To investigate this discrepancy, we tested three alternative SFR-dependent obscuration correction slopes (models 1–3). The slope, which serves as a proxy for dust content through deviations from a one-to-one H$\alpha$–radio SFR relation, is found to evolve with redshift. Specifically, the required slope decreases toward earlier epochs.

The observed trend toward shallower slopes at earlier epochs indicates a lower average dust content at higher redshift. This result is not inconsistent with previous findings of high obscuration at intermediate and high redshifts \citep{perez-gonzalez_spitzer_2005, le_floch_infrared_2005, stach_alma_2018}, which are often attributed to rare, extremely obscured star-forming galaxies that are not represented in local calibrations. The substantial variation observed at high $z$ therefore suggests that more flexible, multi-dimensional calibrations—potentially incorporating galaxy mass, morphology, or spectral energy distribution shape—are required to robustly estimate SFRs across cosmic time. Recent results from the CEERS survey \citep{bail_jwstceers_2024} support this interpretation, linking variations in dust content at high redshift to galaxy structural and stellar population properties, and reinforcing the need for more physically motivated calibrations.

Such an approach holds promise as a practical and physically motivated method for correcting dust attenuation, particularly in cases where direct measurements (such as spectral energy distribution fitting or Balmer decrements) are unavailable. It is shown here that such luminosity dependent corrections for dust would need to decrease in strength with increasing redshift. With further refinement, including a self-consistent and continuous treatment of redshift evolution, a SFR based dust correction method could serve as a valuable tool in the analysis of large survey data and the interpretation of galaxy evolution over cosmic time. 

\begin{acknowledgement}
We would firstly like to thank Dr. Ian Smail for insightful discussions and helpful suggestions that improved this work.

GAMA is a joint European-Australasian project based around a spectroscopic campaign using the Anglo-Australian Telescope. The GAMA input catalogue is based on data taken from the Sloan Digital Sky Survey and the UKIRT Infrared Deep Sky Survey. Complementary imaging of the GAMA
regions is being obtained by a number of independent survey programmes including GALEX MIS, VST KiDS, VISTA VIKING, WISE, Herschel-ATLAS, GMRT and ASKAP providing UV to radio coverage. GAMA is funded by the STFC (UK), the ARC (Australia), the AAO, and the participating institutions. The GAMA website is http://www.gama-survey.org/.  

This scientific work uses data obtained from Inyarrimanha Ilgari Bundara/the Murchison Radio-astronomy Observatory. We acknowledge the Wajarri Yamaji People as the Traditional Owners and native title holders of the Observatory site. The Australian SKA Pathfinder is part of the Australia Telescope National Facility (https://ror.org/05qajvd42) which is managed by CSIRO. Operation of ASKAP is funded by the Australian Government with support from the National Collaborative Research Infrastructure Strategy. ASKAP uses the resources of the Pawsey Supercomputing Centre. Establishment of ASKAP, the Murchison Radio-astronomy Observatory and the Pawsey Supercomputing Centre are initiatives of the Australian Government, with support from the Government of Western Australia and the Science and Industry Endowment Fund.   

This paper includes archived data obtained through the CSIRO ASKAP Science Data Archive, CASDA (http://data.csiro.au).

\end{acknowledgement}


\bibliography{references}

\appendix

\begin{table*}
\centering
\scriptsize
\begin{tabular}{llll|llll}
\hline
Reference & Estimator & Redshift & log($\rho_{\text{SFR}}$)$^{\mathrm{a}}$ &
Reference & Estimator & Redshift & log($\rho_{\text{SFR}}$)$^{\mathrm{a}}$ \\
\hline
Wyder et al. 2005 & UV & $0.01-0.1$ & $-2.02_{-0.02}^{+0.09}$ &
Terao et al. 2022 & H$\alpha$ & $2.1-2.5$ & $-0.54_{-0.70}^{+0.19}$ \\

Schiminovich et al. 2005 & UV & $0.2-0.4$ & $-1.70_{-0.05}^{+0.05}$ &
Stroe et al. 2015 & H$\alpha$ & $0.15-0.25$ & $-2.03_{0.00}^{+0.04}$ \\

 &  & $0.4-0.6$ & $-1.59_{-0.08}^{+0.15}$ &
 & H$\alpha$ & $0.5-0.8$ & $-1.64_{-0.14}^{+0.12}$ \\

 &  & $0.6-0.8$ & $-1.40_{-0.13}^{+0.31}$ &
Drake et al. 2013 & H$\alpha$ & $0.0-0.35$ & $-2.49_{0.00}^{+0.59}$ \\

 &  & $0.8-1.2$ & $-1.45_{-0.13}^{+0.31}$ &
 &  & $0.0-0.5$ & $-2.15_{-0.02}^{+0.02}$ \\

 &  & $0.6-0.8$ & $-1.44_{-0.1}^{+0.1}$ &
Hayes et al. 2010 & H$\alpha$ & $2.18-2.21$ & $-0.87_{-0.24}^{+0.15}$ \\

 &  & $0.8-1.0$ & $-1.19_{-0.08}^{+0.09}$ &
Ly et al. 2007 & H$\alpha$ & $0.07-0.09$ & $-2.07_{0.00}^{+0.00}$ \\

 &  & $1.0-1.2$ & $-1.14_{-0.09}^{+0.09}$ &
 &  & $0.23-0.25$ & $-2.31_{0.00}^{+0.00}$ \\

 &  & $1.2-1.7$ & $-1.15_{-0.08}^{+0.15}$ &
 &  & $0.39-0.41$ & $-1.99_{0.00}^{+0.00}$ \\

 &  & $1.7-2.5$ & $-0.95_{-0.09}^{+0.49}$ &
Yan et al. 1999 & H$\alpha$ & $0.7-1.9$ & $-1.16_{-0.11}^{+0.09}$ \\

 &  & $2.5-3.5$ & $-1.24_{-0.15}^{+0.26}$ &
Moorwood et al. 2000 & H$\alpha$ & $1.3-2.2$ & $-1.12_{0.00}^{+0.00}$ \\

 &  & $3.5-4.5$ & $-1.89_{-0.32}^{+0.22}$ &
Hippelein et al. 2003 & H$\alpha$ & $0.238-0.252$ & $-2.03_{-0.13}^{+0.10}$ \\

Dahlen et al. 2007 & UV & $0.92-1.33$ & $-1.22_{-0.08}^{+0.08}$ &
Glazebrook et al. 2004 & H$\alpha$ & $0.378-0.39$ & $-2.25_{-0.21}^{+0.14}$ \\

 &  & $1.62-1.88$ & $-0.95_{-0.12}^{+0.12}$ &
 &  & $0.452-0.464$ & $-1.27_{-0.14}^{+0.17}$ \\

 &  & $2.08-2.37$ & $-1.07_{-0.09}^{+0.09}$ &
Sanders et al. 2003 & IR & $0.03$ & $-1.92_{-0.03}^{+0.02}$ \\

Reddy et al. 2009 & UV & $1.9-2.7$ & $-0.95_{-0.11}^{+0.09}$ &
Takeuchi et al. 2003 & IR & $0.03$ & $-2.15_{-0.20}^{+0.20}$ \\

 &  & $2.7-3.4$ & $-1.17_{-0.15}^{+0.11}$ &
Magnelli et al. 2011 & IR & $0.40$–$0.70$ & $-1.54_{-0.11}^{+0.22}$ \\

Bouwens et al. 2012 & UV & $3.8$ & $-1.49_{-0.05}^{+0.05}$ &
 &  & $0.70$–$1.00$ & $-1.16_{-0.19}^{+0.15}$ \\

 &  & $4.9$ & $-1.62_{-0.06}^{+0.06}$ &
 &  & $1.00$–$1.30$ & $-1.09_{-0.21}^{+0.27}$ \\

 &  & $5.9$ & $-1.85_{-0.08}^{+0.08}$ &
 &  & $1.30$–$1.80$ & $-1.11_{-0.21}^{+0.17}$ \\

 &  & $7.0$ & $-1.99_{-0.1}^{+0.1}$ &
 &  & $1.80$–$2.30$ & $-1.09_{-0.25}^{+0.21}$ \\

 &  & $7.9$ & $-2.29_{-0.11}^{+0.11}$ &
Magnelli et al. 2013 & IR & $0.40$–$0.70$ & $-1.42_{-0.11}^{+0.08}$ \\

Schenker et al. 2013 & UV & $7.0$ & $-2.20_{-0.11}^{+0.1}$ &
 &  & $0.70$–$1.00$ & $-1.30_{-0.13}^{+0.10}$ \\

 &  & $8.0$ & $-2.41_{-0.14}^{+0.14}$ &
 &  & $1.00$–$1.30$ & $-1.16_{-0.20}^{+0.13}$ \\

Driver et al. 2018 & UV & $0.06$–$0.14$ & $-1.82_{-0.01}^{+0.01}$ &
 &  & $1.30$–$1.80$ & $-1.14_{-0.18}^{+0.13}$ \\

 &  & $0.14$–$0.20$ & $-1.90$ &
 &  & $1.80$–$2.30$ & $-1.00_{-0.15}^{+0.18}$ \\

 &  & $0.20$–$0.28$ & $-1.77$ &
Gruppioni et al. 2013 & IR & $0.00$–$0.30$ & $-1.84_{-0.11}^{+0.09}$ \\

 &  & $0.28$–$0.36$ & $-1.75$ &
 &  & $0.30$–$0.45$ & $-1.62_{-0.04}^{+0.03}$ \\

 &  & $0.36$–$0.45$ & $-1.79_{-0.01}^{+0.01}$ &
 &  & $0.45$–$0.60$ & $-1.52_{-0.05}^{+0.05}$ \\

 &  & $0.45$–$0.56$ & $-1.73_{-0.01}^{+0.01}$ &
 &  & $0.60$–$0.80$ & $-1.34_{-0.06}^{+0.06}$ \\

 &  & $0.56$–$0.68$ & $-1.56$ &
 &  & $0.80$–$1.00$ & $-1.14_{-0.06}^{+0.05}$ \\

 &  & $0.68$–$0.82$ & $-1.42_{-0.01}^{+0.01}$ &
 &  & $1.00$–$1.20$ & $-1.01_{-0.05}^{+0.04}$ \\

 &  & $0.82$–$1.00$ & $-1.29$ &
 &  & $1.20$–$1.70$ & $-1.04_{-0.04}^{+0.04}$ \\

 &  & $1.00$–$1.20$ & $-1.31$ &
 &  & $1.70$–$2.00$ & $-1.06_{-0.03}^{+0.02}$ \\

 &  & $1.20$–$1.45$ & $-1.27$ &
 &  & $2.00$–$2.50$ & $-1.11_{-0.12}^{+0.09}$ \\

 &  & $1.45$–$1.75$ & $-1.17$ &
 &  & $2.50$–$3.00$ & $-1.06_{-0.23}^{+0.15}$ \\

 &  & $1.75$–$2.20$ & $-1.30_{-0.01}^{+0.01}$ &
 &  & $3.00$–$4.20$ & $-1.56_{-0.50}^{+0.23}$ \\

 &  & $2.20$–$2.60$ & $-1.29_{-0.01}^{+0.01}$ &
Novak et al. 2017 & Radio & $0.102-0.4$ & $-1.97_{-0.052}^{+0.047}$ \\

 &  & $2.60$–$3.25$ & $-1.28_{-0.01}^{+0.01}$ &
 &  & $0.401-0.6$ & $-1.89_{-0.050}^{+0.045}$ \\

 &  & $3.25$–$3.75$ & $-1.33_{-0.01}^{+0.01}$ &
 &  & $0.6-0.805$ & $-1.67_{-0.048}^{+0.044}$ \\

 &  & $3.75$–$4.25$ & $-1.42_{-0.04}^{+0.04}$ &
 &  & $0.803-1.0$ & $-1.51_{-0.048}^{+0.044}$ \\

 &  & $4.25$–$5.00$ & $-1.45_{-0.04}^{+0.04}$ &
 &  & $1.0-1.3$ & $-1.51_{-0.049}^{+0.043}$ \\

Covelo-Paz et al. 2024 & H$\alpha$ & $3.8-5.1$ & $-1.38_{-0.12}^{+0.05}$ &
Enia et al. 2022 & Radio & $0.1-0.4$ & $-1.78_{-0.24}^{+0.24}$ \\

 &  & $4.9-5.7$ & $-1.54_{0.00}^{+0.04}$ &
 &  & $0.4-0.7$ & $-1.38_{-0.14}^{+0.14}$ \\

 &  & $5.7-6.6$ & $-1.85_{-0.54}^{+0.11}$ &
 &  & $0.7-1.0$ & $-1.16_{-0.16}^{+0.16}$ \\

Bollo et al. 2023 & H$\alpha$ & $3.86-4.94$ & $-1.19_{-0.16}^{+0.16}$ &
 &  & $1.0-2.0$ & $-0.96_{-0.17}^{+0.17}$ \\

Rinaldi et al. 2023 & H$\alpha$ & $7-8$ & $-2.35_{0.00}^{+0.30}$ &
 &  & $2.0-3.0$ & $-0.99_{-0.22}^{+0.22}$ \\

Sobral et al. 2013 & H$\alpha$ & $0.35-0.45$ & $-2.30_{0.00}^{+0.10}$ &
Condon et al. 2002 & Radio & $0.0-0.1$ & $-2.16_{-0.03}^{+0.03}$ \\

 &  & $0.75-0.95$ & $-1.60_{-0.12}^{+0.07}$ &
Sadler et al. 2002 & Radio & $0.0-0.16$ & $-1.93_{-0.09}^{+0.08}$ \\

 &  & $1.35-1.55$ & $-1.35_{-0.07}^{+0.06}$ &
Serjeant et al. 2002 & Radio & $0.0-0.02$ & $-1.95_{-0.10}^{+0.08}$ \\

 &  & $1.7-2.8$ & $-1.09_{-0.03}^{+0.03}$ &
Machalski et al. 2000 & Radio & $0.0-0.14$ & $-2.12_{-0.10}^{+0.10}$ \\

 &  &  &  &
Haarsma et al. 2000 & Radio & $0.010-0.401$ & $-1.59_{-0.21}^{+0.14}$ \\

 &  &  &  &
 &  & $0.410-0.518$ & $-1.38_{-0.20}^{+0.14}$ \\

 &  &  &  &
 &  & $0.548-0.698$ & $-1.32_{-0.26}^{+0.16}$ \\

 &  &  &  &
 &  & $0.724-0.884$ & $-1.08_{-0.18}^{+0.13}$ \\

 &  &  &  &
 &  & $0.960-4.420$ & $-0.99_{-0.18}^{+0.12}$ \\
\hline
\multicolumn{8}{l}{\small $^{\mathrm{a}}$ Calibrated for Chabrier IMF where needed.}
\end{tabular}
\caption{Measurements of SFR Density, $\rho_{\text{SFR}}$ [M$_{\odot}$ yr$^{-1}$ Mpc$^{-3}$].}
\label{Tab:SFRD_combined}
\end{table*}

\end{document}